\documentclass[12pt]{iopart}
\usepackage{iopams}  
\usepackage{graphicx}
\usepackage{color}
\usepackage{subfigure}
\usepackage[dvipsnames]{xcolor}

\usepackage{braket}

\usepackage{tikz}
\usetikzlibrary{arrows,intersections}
\usepackage{pgfplots}
\usepackage[all]{xy}

\begin{document}

\title[Numerical Coalescence]{Numerical Coalescence of Chaotic Trajectories}

\author{Bruce N. Roth and Michael Wilkinson$^1$}

\address{
$^1$ School of Mathematics and Statistics,
The Open University, Walton Hall, Milton Keynes, MK7 6AA, England\\
}
\vspace{10pt}
\ead{
m.wilkinson@open.ac.uk
}
\begin{indented}
\item December 2019 
\end{indented}

\begin{abstract}
Pairs of numerically computed trajectories of a chaotic system may coalesce because of finite 
arithmetic precision. We analyse an example of this phenomenon, showing that it occurs surprisingly 
frequently. We argue that our model belongs to a universality class of chaotic systems where 
this numerical coincidence effect can be described by mapping it to a first-passage process. 
Our results are applicable to aggregation of small particles in random flows, as well as to 
numerical investigation of chaotic systems. 
\end{abstract}

\maketitle

\section{Introduction}
\label{sec: 1}

When we numerically compute two distinct trajectories of a deterministic system, after a 
certain number of iterations their coordinates may happen to be exactly equal, to machine 
precision. At this point all subsequent computations of the two trajectories yield 
exactly the same values. We might say that the trajectories have undergone \emph{numerical coalescence}. 
This coalescence effect is readily observed in systems with stable dynamics, where nearby 
trajectories approach each other, typically with an exponentially decreasing separation. 
It might, however, be expected that it would be unusual to see this numerical coalescence in 
chaotic systems, where nearby trajectories have an exponentially growing separation 
(the concept of chaos in the theory of dynamical systems is discussed in \cite{Ott,Strogatz}). 
In this paper we analyse numerical coalescence in a simple 
dynamical system which shows a transition to chaos, exploring its dependence upon 
the numerical precision of the calculation, which is denoted by $\delta$. 
In our numerical work we used variable precision arithmetic implemented 
in the \emph{mpmath} package \cite{mpmath} for the Python programming language, and 
comparable results were obtained using the Maple mathematical 
software system \cite{Maple}. These packages allow the number of 
decimal digits, $M$, to be set to an arbitrary integer value. 
If the typical magnitude of the numbers representing the trajectory 
is $\bar x$, we may write 
\begin{equation}
\label{eq: 1.1}
\delta\sim \bar x\, 10^{-M}
\ .
\end{equation}
Numerical experiments showed that coalescence is 
surprisingly frequent. We were able to explain this observation, showing that 
our numerical results are in agreement with a calculation which 
maps the numerical coalescence to a first-passage problem \cite{Red01} for a simple 
stochastic process. We argue that our results are representative of a physically 
significant universality class of chaotic systems.

As well as being relevant to the numerical exploration of complex dynamical systems, the effect 
can also serve as a model for aggregation of small particles in complex flows (reviewed in 
\cite{Pum+16,Gus+16}). If the particles aggregate when their separation falls below a threshold, $\delta$, which 
is analogous to computing trajectories with a finite numerical precision. 

Our explanation of the numerical coalescence phenomenon is related to 
earlier studies \cite{Wil+12,Pradas}, where it was shown that 
trajectories of a chaotic system can display a clustering effect, which is 
characterised by a power-law distribution of their separations. We 
review the relationship between the results in this paper and those earlier 
works in our conclusions, section \ref{sec: 5}.  

\section{An example}
\label{sec: 2}

We start with a simple example illustrating the effect that we wish to explain and analyse.
Consider trajectories of a one-dimensional map
\begin{equation}
\label{eq: 2.1}
x_{n+1}=x_n+F(x_n-\phi_n)
\end{equation}
in the case where $F$ is a function which is continuous and differentiable 
almost everywhere, and which has unit period, $F(x+1)=F(x)$.  
The $\phi_n$ are random numbers, chosen independently at each time step, with a 
distribution which is uniform on $[0,1]$. We also found it convenient to require that the mean value 
of $F(x)$ is equal to zero. Note that, while a single trajectory of (\ref{eq: 2.1}) executes a random walk, 
this equation describes a differentiable dynamical system, so that it is possible to calculate 
the Lyapunov exponent \cite{Ott}. A system with similar properties to (\ref{eq: 2.1}) 
is discussed in some detail in \cite{Wil+03}.

In our study the function $F(x)$ was piecewise 
linear, having a sawtooth form illustrated by figure 1
, with gradients $\pm g$ on intervals 
of length $1/2$, 
\begin{equation}
\label{eq: 2.2}
F(x)= \left\{
\begin{array}{ll}
gx-\frac{g}{4}\quad & 0 \le x \le \frac{1}{2}\\
g(1-x)-\frac{g}{4}\quad & \frac{1}{2}\le x \le 1
\end{array}  \right.
\ .
\end{equation}
With this choice of $F(x)$, an individual trajectory $x_n$ executes a random 
walk for which the variance of the step sizes is $g^2/48$, implying that 
\begin{equation}
\label{eq: 2.3}
\langle x_n\rangle=0
\ ,\ \ \ 
\langle x_n^2\rangle =2{\cal D}n
\ ,\ \ \ 
{\cal D}=\frac{g^2}{96}
\end{equation}
where $\langle \cdot \rangle$ denotes expectation value.

Because the map is periodic, with period unity, if two trajectories $x_n$ and $x'_n$ 
differ by an integer, then $x_n-x'_n$ remains equal to the same integer for all subsequent 
iterations. Note that the map has multiple pre-images when $g>1$.  We shall assume that $g>1$ 
throughout our discussion of the map defined by (\ref{eq: 2.1}) and (\ref{eq: 2.2}).

\begin{figure}[h]
\label{fig: 1}
\setlength{\unitlength}{5mm}
\begin{picture}(0,9.5)(0,-1)
  \put(1,3){\vector(1,0){25}}
  \put(10.5,-0.5){\vector(0,1){8.5}}
  \put(10.5,0){\line(1,2){3}}
  \put(16.5,0){\line(-1,2){3}}
  \put(16.5,3){\line(0,-1){0.3}}
  \put(16.3,2){$1$}
  \put(16.5,0){\line(1,2){3}}
  \put(22.5,0){\line(-1,2){3}}
  \put(22.5,3){\line(0,-1){0.3}}
  \put(22.3,2){$2$}
  \put(4.5,0){\line(1,2){3}}
  \put(10.5,0){\line(-1,2){3}}
  \put(4.5,3){\line(0,-1){0.3}}
  \put(4,2){$-1$}
  \put(22.5,0){\line(1,2){3}}
  \put(4.5,0){\line(-1,2){3}}
  \put(8.2,7.5){$F(x)$}
  \put(26,2){$x$}
\end{picture}
\caption{Sawtooth function used in (\ref{eq: 2.1})}
\end{figure}
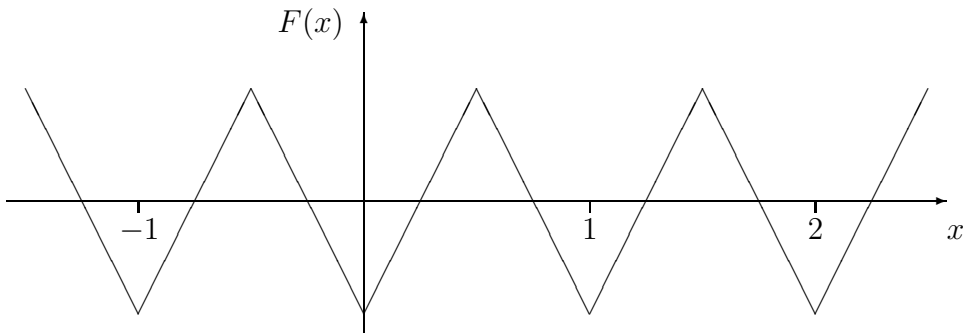

In section \ref{sec: 3} we demonstrate that the map is chaotic (because the 
Lyapunov exponent, $\lambda$, is greater than zero \cite{Ott})  for $g>\sqrt{2}$. When $\lambda>0$, nearby 
trajectories should separate exponentially \cite{Ott}: if $\Delta x_n$ is the separation 
of two trajectories after $n$ iterations, then 
\begin{equation}
\label{ eq: 2.4}
\lambda_n=\frac{1}{n}\ln\left(\frac{\Delta x_n}{\Delta x_0}\right)
\end{equation}
approaches the Lyapunov exponent $\lambda$ as $n\to \infty$, provided $\Delta x_0$ is sufficiently small 
that $|\Delta x_n|\ll 1$. When $\lambda<0$, trajectories converge to 
the same point (if their initial separation is small) or else to integer separations 
due to the periodic nature of the one-dimensional map. 
When $\lambda>0$, it might be expected that a very small initial separation of trajectories 
eventually grows to be of order unity, and the similarity of (\ref{eq: 2.1}) to the 
equation of a random walk leads one to expect that subsequent growth of separation 
is diffusive, $\Delta x\sim \sqrt{n}$. 

However, numerical investigations of (\ref{eq: 2.1}) show a different 
behaviour. Table 1 
lists the separations for pairs of trajectories with $x_0$ random and 
an initial separation of $\Delta x_0=0.25$ after $n=2.5\times 10^5$ iterations of the map, for a 
range of values of $g$.  Here we used an arithmetic precision which was defined by 
specifying $M=10$ decimal places.
The numerically computed separations of 
the two trajectories, $x_n$ and $x_n'$, are either 
exactly zero or exactly integers, for two 
values of $g$ which are in excess of the threshold 
for chaos, at $g=\sqrt{2}=1.41421\ldots$. 

\begin{table}
\begin{tabular}{c c c c}
\label{tab: 1}
$g$ & $x_n$ & $x_n'$ & $\Delta x_n$\\
\hline
1.3   & -94.67387172 &-93.67387172 &-1.0\\
1.35 & 41.57990309 &44.57990309 &-3.0\\
1.4   & 41.41722753 &41.41722753 &0.0\\
1.45 & 84.31604915 &84.31604915 & 0.0\\
1.5   & 46.59522684 & 46.59522684 & 0.0\\
1.55 & 68.70661011  &172.6956409 & -103.9890308\\
1.6   & -60.55216735 &  -147.5451987 & 86.9930313\\
1.65 & -48.66364209 &  155.3371666 & -204.0008087\\
1.7   & -154.9828462 & -122.6170971 & -32.36574912\\
\end{tabular}
\caption{Two trajectories of a realisation of (\ref{eq: 2.1}) after $n=2.5\times 10^5$ 
iterations, computed with $M=10$ digit arithmetic. One of the initial conditions was random in $[0,1]$, and the 
initial separation was $\Delta x_0=0.25$ in all cases.}
\end{table}

It is easy to understand why this numerical coalescence effect is possible for
values of $g$ beyond the threshold for chaos. 
When $\lambda>0$, it is still possible for trajectories to coalesce 
if two trajectories happen to reach \emph{exactly} the same 
point (this is possible because the calculation uses finite precision 
arithmetic, and because points may have multiple pre-images).
However, we might expect that this occurrence is rare. 
Above the threshold for chaos, nearby points are separating exponentially. 
If these separations were to fill the unit interval with constant density, there would be  
a probability $P\sim \delta$ to coalesce at each iteration of the map. 
If the probability to coalesce were $\delta$ upon each iteration, then after $n$ iterations 
the probability $P_{\rm c}$ for the paths to have undergone coalescence would be 
\begin{equation}
\label{eq: 2.5}
P_{\rm c}\sim n\bar x\,10^{-M}
\ .
\end{equation}
In our numerical investigation illustrated by table 1, 
$n=2.5\times 10^5$, implying (from equation (\ref{eq: 2.3}), using 
$g=2$ as a representative value) that $\bar x=\sqrt{2{\cal D}n}\approx 100$.  
This would give an estimate for the probability for coalescence of  
$P_{\rm c}\sim 2.5\times 10^5 \times 100 \times 10^{-10}=2.5\times 10^{-3}$.  
According to this estimate, the results displayed in table 1 
 appear to be highly unlikely. 
The remainder of this paper will explain and quantify the effect illustrated in table 1, 
(section \ref{sec: 3}), and discuss the extent to which it is a manifestation 
of a universal phenomenon (section \ref{sec: 4}). 

\section{Coalescence in finite precision arithmetic}
\label{sec: 3}

\subsection{Numerical investigation}
\label{sec: 3.1}

When two trajectories, started from randomly chosen initial conditions, reach 
exactly the same coordinate (or else 
two coordinates with an exactly integer separation), their values 
remain locked together for all subsequent iterations. If the arithmetic 
of a numerical iteration of the map has a finite precision, say $M$ 
decimal places, then this phenomenon of \emph{numerical coalescence} 
provides an explanation for the effect illustrated in table 1
. The principal difficulty is to explain why the effect happens so much more frequently 
than the simple estimate, equation (\ref{eq: 2.5}).

We can take two trajectories and determine the \lq time' $N$ (that is, number 
of iterations) for them to coalesce. This will be different for different 
initial conditions. We should therefore look at the probability $P(N)$, 
or else at statistics such as the moments $\langle N^j\rangle$. 
The simplest statistic is just the mean time for coalescence, $\langle N\rangle$, 
and the remainder of this section is concerned with estimating this quantity.

We investigated the mean number of iterations for coalescence of trajectories $\langle N\rangle$
as a function of $g$ and of the number of decimal digits, $M$. The results are presented in table 2. 
The initial separation was $\Delta x_0=0.1$ in all cases, and for all of the data points 
we averaged over $1000$ realisations, with initial conditions distributed randomly with uniform density on $[0,1]$. 

Note that most of these values of $\langle N\rangle$ are sufficiently small that 
$\bar x\sim \sqrt{2{\cal D}\langle N\rangle}$ is not a very large number: for 
example at $g=1.4$ and $M=20$, we found $\langle N\rangle\approx 2500$, implying 
that $\bar x\approx 10$, which indicates that only one of the $M=20$ digits is required to store the 
integer part of $x_n$ leaving $19$ digits after the decimal point. 
For this reason we our analysis of these data will make the simplifying assumption that 
the precision of the calculation, $\delta$, is given by $\delta=10^{-M}$, 
where $M$ is the number of digits.

\begin{table}
\begin{tabular}{c  |c c c c c c c}
 & & & & $\langle N\rangle$ & & & \\
$g$ & $M=7$& $M=10$ & $M=15$ & $M=20$ & $M=30$ & $M=40$ & $M=50$\\
\hline
1.25  &  58.0   &  81.8 &  120   &  159 &  237  &   318  &  396 \\
1.3    & 83.0    & 120  & 177  &   243  & 363  &   481  &  612 \\
1.32  & 101  &  141  & 217 &    295  & 444  &   599  &  761 \\
1.34  & 119  &  180  & 273  &   371  & 566   &  773  &  957 \\
1.36  & 151  &  230  & 371 &    502  & 785   & $1.08\times 10^3$  & $1.33\times 10^3$ \\
1.38  & 193  &  323  & 523 &    739  & $1.17\times 10^3$  & $1.66\times 10^3$  & $2.11\times 10^3$ \\
1.39  & 227  &  392  & 668 &    957  & $1.61\times 10^3$  & $2.24\times 10^3$  &  $2.92\times 10^3$ \\
1.4    & 272  &  467  & 883 &   $1.42\times 10^3$  & $2.48\times 10^3$  & $3.54\times 10^3$  & $4.54\times 10^3$ \\
1.41  & 327  &  611  & $1.29\times 10^3$ &  $2.10\times 10^3$  & $4.41\times 10^3$  & $6.96\times 10^3$ &  $9.75\times 10^3$ \\
1.42  & 413  &  828  & $1.91\times 10^3$ &  $3.73\times 10^3$  & $1.04\times 10^4$ & $2.00\times 10^3$ & $3.02\times 10^4$ \\
1.43  & 476  & $1.07\times 10^3$ & $3.25\times 10^3$ &  $8.34\times 10^3$ & $2.82\times 10^4$ &                    &           \\
1.44  & 604  & $1.68\times 10^3$ &  $5.88\times 10^3$ & $1.97\times 10^4$ &                  &                    &           \\
1.45  & 815  & $2.42\times 10^3$ & $1.19\times 10^4$ & $3.09\times 10^4$ &                 &                    &            \\
1.46  & $1.07\times 10^3$ & $3.93\times 10^3$ & $2.41\times 10^4$ &                  &                 &                   &            \\
1.47  & $1.50\times 10^3$ &  $6.25\times 10^3$ & $5.75\times 10^4$ &                 &                 &                   &             \\
1.48  & $1.89\times 10^3$ & $1.08\times 10^4$ & $7.57\times 10^4$ &                &                 &                   &             \\
1.49  & $2.76\times 10^3$ & $1.85\times 10^4$  & $5.73\times 10^4$ &               &                 &                   &              \\
\end{tabular}
\label{tab: 2}
\caption{Tabulation of the mean number of iterations before coalescence, $\langle N\rangle$, 
determined numerically for different values of the coefficient $g$ and of the number of digits, $M$. 
The Lyapunov exponent $\lambda$, given by the first line of equations (\ref{eq: 3.2}), is 
positive when $g>\sqrt{2}=1.41\ldots$.}
\end{table}

\subsection{Theory: relation to mean first-passage time}
\label{sec: 3.2}

Next we consider how to analyse the results in table 2. 
Because the map (\ref{eq: 2.1}) has unit period, two trajectories which have 
an exactly integer separation maintain the same separation for all subsequent iterations.
In this sense, integer separations of trajectories are equivalent to a 
zero separation, and accordingly we define $\Delta x_n$ as 
the magnitude of the separation of two trajectories modulo unity. In order to facilitate the calculation 
of $\langle N\rangle$, the separation of two trajectories, $\Delta x$, is transformed into a 
logarithmic variable, 
\begin{equation}
\label{eq: 3.0}
Z\equiv -\ln\,|\Delta x|
\ .
\end{equation}
If trajectories were computed with arbitrary precision, 
the variable $Z$ would occupy the interval $Z\in[0,\infty)$. When $Z$ is large, the 
separation of two trajectories is very small, and (at points where the derivative of the 
map exists) the iteration of $Z$ is described by linearisation of the map, so that
\begin{equation}
\label{eq: 3.1}
Z_{n+1}=Z_n-\ln\,\left\vert\frac{\partial x_{n+1}}{\partial x_n}\right\vert
\ .
\end{equation}
For the map (\ref{eq: 2.1}), the dynamics of (\ref{eq: 3.1}) is Markovian, with 
displacements $\Delta Z_\pm=-\ln (1\pm g)$ occurring with random choices of the sign, 
having probabilities $P_\pm=1/2$. 
Because (\ref{eq: 3.1}) is the equation of a random walk, 
over many iterations the motion of $Z$ can be modelled by an 
advection-diffusion equation, with a drift velocity $v$ and a diffusion coefficient $D$
(see \cite{vKa81} for a discussion of the relationships between random walks or Langevin processes 
and diffusion or Fokker-Planck equations). 
Note that, from the definition of the Lyapunov exponent $\lambda$ \cite{Ott}, we have $\lambda=-v$. 
The values of $v$ and $D$ are determined from the statistics of $\Delta Z$: 
if time is measured by the number of iterations, the 
drift velocity $v$ and diffusion coefficient $D$ are obtained from 
\begin{eqnarray}
\label{eq: 3.2}
&&v=-\lambda=\langle \Delta Z\rangle =-\frac{1}{2}\left[ \ln(g+1)+\ln(g-1)\right]
\ ,
\nonumber \\
&&\langle \Delta Z^2\rangle=\frac{1}{2}\left[\ln(g+1)\right]^2+\frac{1}{2}\left[\ln(g-1)\right]^2
\ ,
\nonumber \\
&&D=\frac{1}{2}\left[\langle \Delta Z^2\rangle-\langle \Delta Z\rangle^2\right]
\ .
\end{eqnarray}
The first of these relations shows that $\lambda=\ln\sqrt{g^2-1}$, which is positive when $g>\sqrt{2}$. 

When $\Delta x$ is small, implying that $Z$ is large, the behaviour of $Z$ is determined by the linearised equation 
of motion, leading to (\ref{eq: 3.1}). In the vicinity of $Z=0$, however, $Z$ has complex dynamics 
which could, in principle, be determined from the equation of motion of $x$, (that is, from equation (\ref{eq: 2.1})).  
However we can observe that the representative point $Z$ does not 
pass $Z=0$, and eventually will leave the region close to $Z=0$. This implies that the diffusion process can be 
modelled as having a reflecting barrier at $Z=0$. If we take account of the coalescence 
of trajectories due to finite-precision representation of arithmetic, this implies that the separation 
of trajectories becomes zero when $\Delta x=\delta=10^{-M}$ where $\delta$ is the floating point precision. 
At this point, the diffusive model 
ceases to be appropriate. We can represent this by introducing an absorbing barrier at a position
\begin{equation}
\label{eq: 3.3}
Z_0=-\ln \delta=M\ln 10
\ .
\end{equation}
The mean number of iterations before coalescence, 
$\langle N\rangle$, is the same as the mean time for first contact with the absorbing barrier 
in the diffusion process. This quantity is known as the \emph{mean first-passage time}. 

\subsection{Calculation of mean time to coalescence}
\label{sec: 3.3}

There is an extensive literature on the mean first-passage time for diffusive processes 
\cite{Red01}. We shall use a standard formula for the mean first-passage time for the Langevin equation 
\begin{equation}
\label{eq: 3.4}
\dot Z=-\frac{{\rm d}V}{{\rm d}Z}+\sqrt{2D}\eta(t)
\end{equation}
where $V(Z)$ is a potential and $\eta(t)$ is white noise. We assume that there is 
a reflecting barrier at $Z_1$, absorbing barrier at $Z_0$, 
and particles initially released from $Z_{\rm i}$. The mean first-passage time 
is (see \cite{Lif+62,Zwanzig1988})
\begin{equation}
\label{eq: 3.5}
\langle T\rangle=\frac{1}{D}\int_{Z_{\rm i}}^{Z_0}{\rm d}x\ \exp[V(x)/D]\int_{Z_1}^x {\rm d}y\ \exp[-V(y)/D]
\ .
\end{equation}
As $Z$ approaches $\infty$, the mean velocity $v$ is becomes independent of $Z$, and is 
equal to $-\lambda$, where $\lambda$ is the Lyapunov exponent. The dynamics of $Z$ is also subject to fluctuations 
about this mean motion (which are quantified by the diffusion coefficient $D$). If the potential is $V(x)=-vx$, 
then the drift velocity, $-V'(x)$, is a constant, $v$, as required. The true boundary condition at $Z=0$ could be 
modelled more faithfully by introducing a delay time, but we shall assume that using a reflecting boundary at 
$Z=0$ is sufficient. We could, in principle, average over the initial conditions by averaging $Z_i$ over 
the steady-state distribution of $Z$, but again we adopt a simplifying assumption, assuming all of the 
representative points are injected at $Z_i=0$ (we expect that this approximation will cause our calculation 
to overestimate the time taken to reach the absorbing barrier). Defining 
\begin{equation}
\label{eq: 3.6}
\alpha=\frac{v}{D}
\end{equation}
and setting $V(x)=-vx$ in equation (\ref{eq: 3.5}), we obtain
\begin{equation}
\label{eq: 3.7}
\langle T \rangle =\frac{1}{D}\int_0^{Z_0}{\rm d}x\ \exp(-\alpha x)\int_0^x {\rm d}y\ \exp(\alpha y)
\ .
\end{equation}
Evaluating the integrals gives
\begin{equation}
\label{eq: 3.8}
\langle T\rangle=\frac{1}{v\alpha}\left[\exp(-\alpha Z_0)-1\right]+\frac{Z_0}{v}
\ .
\end{equation}
Noting that the expected number of iterations for path coalesce is 
$\langle N\rangle=\langle T\rangle$, this can be written in the form 
\begin{equation}
\label{eq: 3.9}
\langle N\rangle=\frac{Z_0^2}{D}f(X)
\end{equation}
where 
\begin{equation}
\label{eq: 3.10}
X=\frac{\lambda Z_0}{D}
\end{equation}
and 
\begin{equation}
\label{eq: 3.11}
f(X)=\frac{\exp(X)-1-X}{X^2}
\ .
\end{equation}
Because this approximation assumes $|Z_0|\gg 1$, the predicted values of $\langle N\rangle$ are 
un-observably large or small unless $g$ is close enough
to $g_0=\sqrt{2}$ that we can make the following two approximations:
\begin{eqnarray}
\label{eq: 3.12}
&&D\approx D_0=\frac{[\ln\,(\sqrt{2}-1)]^2+[\ln\,(\sqrt{2}+1)]^2}{4}\approx 0.38841
\ ,
\nonumber \\
&&v\approx \frac{\partial \lambda}{\partial g}\bigg\vert_{g=g_0}(g-g_0)=\sqrt{2}(g-g_0)
\end{eqnarray}
where $g_0=\sqrt{2}$ is the parameter value for the threshold of chaos, and $D_0$ is the value 
of the diffusion coefficient at $g_0$.   
Using these approximations we find 
\begin{eqnarray}
\label{eq: 3.13}
&&\langle N\rangle \sim \frac{(\ln 10)^2}{D_0}\,M^2\, f(X)\approx 13.65\times M^2\, f(X)
\nonumber \\
&&X=\frac{\sqrt{2}\ln\,10}{D_0}(g-\sqrt{2})\approx 8.384\times  M\,(g-\sqrt{2})
\ .
\end{eqnarray}
The function $f(X)$ is positive for all real $X$. It implies the following limiting behaviours: 
\begin{eqnarray}
\label{eq: 3.14}
\lim_{\lambda\to -\infty}\langle N\rangle \sim \frac{-Z_0}{\lambda}
\nonumber \\
\lim_{\lambda\to +\infty}\langle N\rangle \sim \frac{Z_0^2}{D}\frac{\exp(X)}{X^2}
\nonumber \\
\lim_{\lambda\to 0}\langle N\rangle \sim \frac {Z_0^2}{2D}
\ .
\end{eqnarray}
The first of these is what would be expected from the definition of the Lyapunov exponent. 
The value of $\langle N\rangle$ should not exceed $\delta^{-1}=10^{M}$, so that 
exponential growth in the limit as $\lambda \to \infty$ which is predicted by (\ref{eq: 3.14}) 
will only be correct when $g-g_0$ is sufficiently small. 

Equations (\ref{eq: 3.13}) are expected to give an asymptotic approximation to 
$\langle N\rangle$, which is valid when $M\gg1$ and $g-\sqrt{2}\ll 1$. 
We compared this theory against the data tabulated in table 2 
by testing whether it would collapse onto a single curve, representing the function $f(X)$. 
Because of the exponential growth of $f(X)$ for positive $X$, it is more convenient 
to graph $\ln\,f(X)$. Accordingly, in order to test the prediction contained in 
equations (\ref{eq: 3.9}) to (\ref{eq: 3.11}), we made a plot of 
\begin{equation}
\label{eq: 3.15}
Y=\ln \left(\frac{\langle N\rangle}{13.65\times M^2}\right)
\end{equation}
as a function of $X$, as defined by equation (\ref{eq: 3.13}). In figure 2 
we display our plot of $Y$ versus $X$ for the data in table 2. 
We use different point styles (and colours, online) to distinguish the 
values of $M$, and included some additional data for more densely sampled 
values of the coefficient $g$. The points with different colours all collapse onto the same curve, 
which is well approximated by the function $Y(X)=\ln f(X)$, plotted as a solid curve. 
This validates the theory described by (\ref{eq: 3.9}), (\ref{eq: 3.10}) and (\ref{eq: 3.11}) as a description of the 
finite-precision path-coalescence effect for map (\ref{eq: 2.1}) and (\ref{eq: 2.2}). 
\begin{figure}[h]
\label{fig: 2}
\begin{center}
\includegraphics[width=0.85\textwidth]{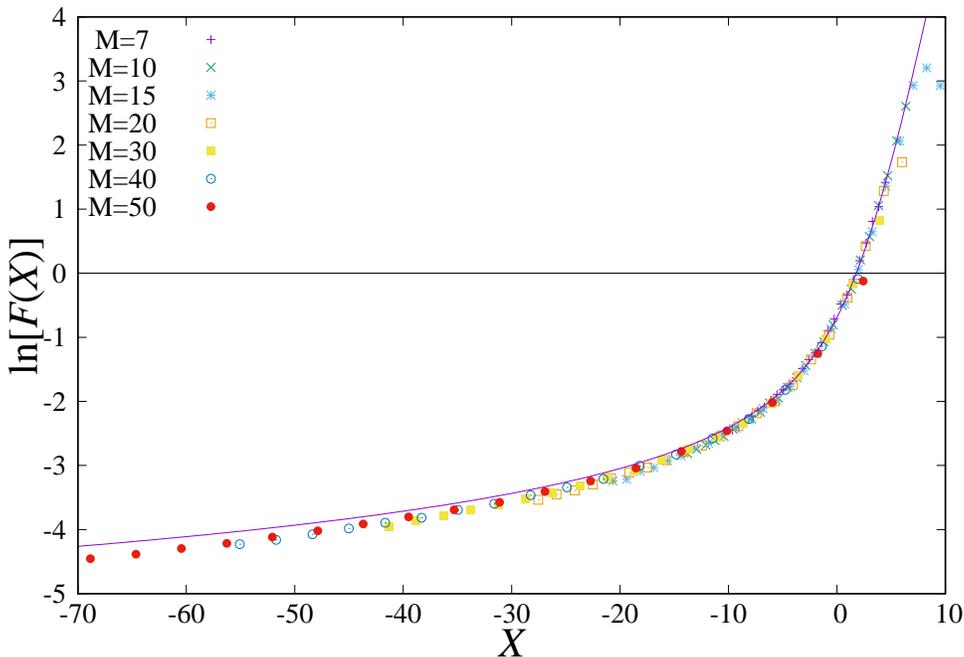}
\end{center}
\caption{Plot  $Y$ against $X$ 
for g from 1.25 to 1.49. Data points for different numbers of digits ($M=7,10,15,20,30,40,50$), 
plotted with different symbols according to the legend on the figure, collapse onto the function 
$\ln\left(\frac{\exp(X)-X-1}{X^2}\right)$ (bold curve).}
\end{figure}

\section{Generalisation to other systems}
\label{sec: 4}

The theory for numerical coalescence which was presented
in section \ref{sec: 3} is applicable to a broad class of dynamical systems, and in this 
sense it may be said to be \lq universal'.

Consider a differentiable map in $d$ dimensions, with a control parameter 
$g$. We denote phase points on a trajectory by $\mbox{\boldmath$x$}_n$, and the small 
separation of two trajectories at iteration $n$ is denoted by $\delta \mbox{\boldmath$x$}_n$. 
If we define $\delta r_n=|\delta \mbox{\boldmath$x$}_n|$ and $Z_n=-\ln\,\delta r_n$, then when 
$\delta r_n$ is small, $\Delta Z_n=Z_n-Z_{n-1}$ depends upon the position $\mbox{\boldmath$x$}_{n-1}$ 
and upon the ratios of the components of $\delta \mbox{\boldmath$x$}_{n-1}$, but the probability 
distribution of $\Delta Z_n$ becomes independent of $Z_n$ as $Z_n\to \infty$. 
Numerical coalescence occurs when $Z_n$ becomes sufficiently large, so that the numerical coalescence 
process in $d$ dimensions can be formulated as a first-passage problem in one dimension. This one-dimensional 
problem can be analysed by the approach used in section \ref{sec: 3}. 

We assume that the dynamical system is either autonomous, or else driven by a process which is at least 
statistically stationary in time. The sequence of values $\Delta Z_n$ has mean and correlation
statistics
\begin{eqnarray}
\label{eq: 4.1}
&&v=\langle \Delta Z_n\rangle
\nonumber \\
&&C_{nm}=\langle \Delta Z_n\Delta Z_m\rangle -v^2
\ .
\end{eqnarray}
If the $\Delta Z_n$ can be regarded as random variables, then the probability density function $P$  
of $\Delta Z$ at iteration $n$ can be modelled by an advection-diffusion equation, where 
$n$ identified with time the $t$. The drift velocity is $v$ and the diffusion coefficient $D$ is 
\begin{equation}
\label{eq: 4.3}
D=\frac{1}{2}\sum_{m=-\infty}^\infty C_{nm}
\ .
\end{equation}
Note that $-v$ is equal to the Lyapunov exponent. 

Both the Lyapunov exponent and the diffusion coefficient are functions of the 
control parameter $g$. Our theory is applicable in the vicinity of the transition to 
chaos, when $g\approx g_0$ defined by $v(g_0)=0$. We also define 
\begin{equation}
\label{eq: 4.4}
v'_0=\frac{{\rm d}v}{{\rm d}g}\bigg\vert_{g=g_0}
\ ,\ \ \ 
D_0=D(g_0)
\ .
\end{equation}
Provided $v'_0$ and $D_0$ are non-zero, we expect that 
the mean time for coalescence is given by equation (\ref{eq: 3.9}), where the parameter $X$ 
is given by
\begin{equation}
\label{eq: 4.5}
X=\frac{g'_0\ln\,10}{D_0}(g-g_0)
\ .
\end{equation}

The theory presented above is expected to be applicable when 
$D_0$ and $v'_0$ are both non-zero. This is expected to hold when the specification of the 
dynamical system includes noise, so that the quantities $\Delta Z_n$ are random numbers. 
The system that we considered in our numerical example, defined by equations 
(\ref{eq: 2.1}) and (\ref{eq: 2.2}), is a concrete example. 

In the case of an autonomous dynamical system, however, the assumptions leading to 
equations (\ref{eq: 3.9}) to (\ref{eq: 3.11}) may not hold. 
In the case where the transition to chaos arises because a stable periodic orbit 
either disappears, or else undergoes a bifurcation, the $\Delta Z_n$ form a periodic sequence 
when $g$ is on the stable side of the transition. In this case the diffusion coefficient $D$ is equal
to zero in the stable phase, and we must have $D_0=0$, so that the theory is not applicable.

\section{Conclusions}
\label{sec: 5}

We have exhibited an example of a chaotic system where trajectories coalesce at a surprisingly 
high rate in the vicinity of the threshold of chaos due to arithmetic truncation errors. 
The effect was explained by a transformation of the separation of two trajectories to a logarithmic 
variable, which leads to analogy with a first-passage problem for a diffusive process. 
This effect is a consequence of transient convergence of chaotic trajectories, which was 
previously analysed in \cite{Wil+12} and \cite{Pradas}. Our analysis of the numerical 
coalescence phenomenon is based upon a principle that was used in those earlier works, 
namely that the linearised equation of motion is mapped to an advection-diffusion equation 
if we make a logarithmic transformation, equation (\ref{eq: 3.1}) in this present work. 
Here we have shown that the numerical coalescence effect may be understood using a 
surprisingly simple treatment of the associated first-passage problem. 

Our analysis is also relevant to understanding the aggregation of particles in complex flows 
where the particles aggregate when their separation is equal to $\delta$. In particular, our theory 
for $\langle N\rangle$ describes the mean time for collision of particles which are advected by a 
flow which has a Lyapunov exponent which is close to zero.

The theory that we have presented is valid when the stability factors $\Delta Z_n$ have 
random or pseudo-random fluctuations. This includes almost all real-world applications 
of chaotic dynamics, where noise will be present. In section \ref{sec: 4} we pointed out 
that the effect will may be absent in systems where the transition to chaos arises 
from perturbation of a periodic orbit. This is consistent with a viewpoint that low-dimensional 
autonomous dynamical systems may not be a good models for physical applications of 
chaos.


\section*{References}

\begin{thebibliography}{99}

\bibitem{Ott}
Ott, E, 2002, {\sl Chaos in Dynamical Systems}, Cambridge.
%
\bibitem{Strogatz}
Strogatz, SH, 2018, 
{\sl Nonlinear Dynamics and Chaos with 
Applications to Physics, Biology, Chemistry, and Engineering}, 2nd edn., 
CRC Press, Boca Raton.
%
\bibitem{mpmath}
Johansson, F., 2013, {\sl mpmath: a Python library for arbitrary-precision floating-point arithmetic}, by 
Fredrik Johansson and others, {\tt http://mpmath.org/}.
%
\bibitem{Maple}
{\tt https://www.maplesoft.com}
%
\bibitem{Red01}
Redner, S., 2001, {\sl A guide to first-passage processes}, Cambridge, University press, ISBN 0-521-65248-0. 
%
\bibitem{Pum+16}
Pumir, A. and Wilkinson, M., 2016, 
{\sl Collisional Aggregation due to Turbulence,}
{\it Ann. Rev. Cond. Matter Phys.}, {\bf 7}, 141-70.
%
\bibitem{Gus+16}
Gustavsson, K and Mehlig, B, 2016,
{\sl Statistical models for spatial patterns of heavy particles in turbulence}, 
{\it Adv. Phys.},  {\bf 65}, 1-57.
%
\bibitem{Wil+12}
Wilkinson, M., Mehlig, B., Gustavsson, K. and Werner, E.,
2012, {\sl Clustering of Exponentially Separating Trajectories},
{\it Eur. Phys. J. B.}, {\bf 85}, 18.
%
\bibitem{Pradas} 
Pradas, M,  Pumir, A., Huber, G and  Wilkinson, M., 
2017, {\sl Convergent chaos}, {\it J. Phys. A: Math. Theor.}, {\bf 50}, 275101. 
%
\bibitem{Wil+03}
Wilkinson, M. and Mehlig, B.,
2003, {\sl The Path-Coalescence Transition and its Applications}, 
{\it Phys. Rev. E}, {\bf 68}, 040101.
%
\bibitem{vKa81}
van Kampen, NG, 1981, 
{\sl Stochastic processes in physics and chemistry}, North Holland, (3rd edn., 2007, ISBN 0-444-89349-0).
%
\bibitem{Lif+62}
Lifson, S. and Jackson, J.L., 1962, 
{\sl On self-diffusion of ions in a polyelectrolyte solution},
{\it J. Chem. Phys.}, {\bf 36}, 2410-14.
%
\bibitem{Zwanzig1988}
Zwanzig, R., 
1988, {\sl Diffusion in a rough potential}, 
{\it PNAS}, {\bf 85}, 2029-30.



\end{thebibliography}
\end{document}